# National mapping and estimation of forest area by dominant tree species using Sentinel-2 data

Johannes Breidenbach [1], Lars T. Waser [2], Misganu Debella-Gilo [1], Johannes Schumacher [1], Johannes Rahlf [1], Marius Hauglin [1], Stefano Puliti [1], Rasmus Astrup [1]

[1] NIBIO, Norwegian Institute of Bioeconomy Research, National Forest Inventory, NO-1400 Ås, Norway

[2] WSL, Swiss Federal Institute for Forest, Snow and Landscape Research, CH-8903 Birmensdorf, Switzerland

## 1   Abstract

Nation-wide Sentinel-2 mosaics were used with National Forest Inventory (NFI) plot data for modelling and subsequent mapping of spruce-, pine- and deciduous-dominated forest types in Norway at a 16m × 16m resolution. The accuracies of the best model ranged between 74% for spruce and 87% for deciduous forest. An overall accuracy of 90% was found on stand level using independent data from more than 42.000 stands. Errors mostly resulting from a forest mask reduced the model accuracies by approximately 10%. The produced models were subsequently used to generate model-assisted (MA) and post stratified (PS) estimates of species-specific forest area. At the national level, efficiencies of the estimates increased by 20% to 50% for MA and up to 90% PS. Greater minimum number of observations constrained the use of PS. For MA estimates of municipalities, efficiencies improved by up to a factor of 8 but were sometimes also below 1. PS estimates were always equally as or more precise than direct and MA estimates but were applicable in fewer municipalities. The tree species prediction map is part of the Norwegian forest resource map and is used, among others, to improve maps of other variables of interest such as timber volume and biomass.

## 2   Introduction

Information on the spatial distribution of tree species is required by the forestry sector and beyond, in particular in a changing climate (Fichtner et al. 2018), for sustainable forest management (Gamfeldt et al. 2013, Haara et al. 2019), climate-change adaptation and mitigation techniques (Hof et al. 2017), and biodiversity assessment (Barbier et al. 2008). Although national forest inventories allow for precise estimates of tree species proportions on regional to national levels, large-scale and fine resolution spatially-explicit information beyond these estimates often is missing outside areas with detailed forest management plans (Barrett et al. 2016). Thus, land use management in general and forest management in particular will benefit from accurate and consistent tree species maps at a national level. Conversely, maps or remotely-sensed products can help improve estimates in forest inventories (McRoberts 2010, McRoberts et al. 2016, Chirici et al. 2020).

Field mapping surveys tend to be costly, in particular over large areas, and are not always feasible due to lack of accessibility for example in steep terrain (Martin et al. 1998). Therefore, remote sensing technologies have frequently been used instead. Still, much of the actual species mapping in forest inventories is done based on manual visual inspections of areal images. Mapping tree species at a national level can nonetheless be a daunting task and most studies have been carried out under optimized conditions, i.e. case studies, rather than countrywide (Fassnacht et al. 2016, Waser et al. 2017).

Over the last 40 years, advances in remote sensing technologies have enabled the classification of tree species using several sensor types. Thus, the number of studies focusing on tree species classification has increased continuously during the same period with varying degrees of success. Most studies have been conducted in boreal forests (North America and Nordic countries) and mixed temperate forest (Europe), and during the last decade also in the tropics. An extensive list of the sensors and methods used to classify tree species is given in the review paper of Fassnacht et al. (2016). Mainly four groups of remote sensing data have been used, all of which are well suited to classifying tree species with high accuracy. These groups are: (1) *Hyperspectral and imaging spectroscopy*, which provides the broadest range of spectral information (e.g. Dalponte et al. 2012, Fricker et al. 2019); (2) *multispectral airborne* (e.g. Waser et al. 2011, Engler et al. 2013) with high spatial but limited spectral resolution, and passive *spaceborne sensors*, which provide a high spectral and relatively high spatial resolution (e.g. Waser et al. 2014, Immitzer et al. 2016, Persson et al. 2018); (3) *lidar*, which provides 3D point clouds (e.g. Hovi et al. 2016, Axelsson et al. 2018); and (4) a *combination of these groups*, mainly involving lidar (e.g. Ghosh et al. 2014, Liu et al. 2018).

The nationwide mapping of tree species based on satellite-based optical remote sensing data is mainly limited by cloud cover. However, thanks to the short revisit period of Sentinel-2 images (i.e. < 3 days in Norway) it is now increasingly feasible to produce cloud-free mosaics even within short periods of time. This not only allows to produce wall-to-wall yearly mosaics to monitor inter-year changes but also to obtain cloud-free intra-year time series that can be used to infer information on the phenology of different tree species. Several studies reveal that tree species predictions improve when images of several seasons are used (Fassnacht et al. 2016, Immitzer et al. 2016, Immitzer et al. 2019) due to phenological differences among different tree species.

Two approaches have been established to map tree species from optical sensors. First, the deterministic approach of modeling the direct response of different tree species to incoming electromagnetic radiation. However, while spectral signatures of most tree species are becoming better studied (Fassnacht et al. 2016, Hovi et al. 2016), they are variable given age and location, and may not be reproducible with operational sensors due to the mix of different species or land-uses in one pixel (Cochrane 2000, Durgante et al. 2013). Second, the empirical (data-driven) approach is an alternative and is based on the relationship between remotely sensed reflectance data and species information from reference observations. Numerous studies used the empirical approach with varying degrees of success (e.g. Immitzer et al. 2016, Holloway and Mengersen 2018, Wessel et al. 2018).

The mapping of forest characteristics by a combination of national forest inventory (NFI) and satellite data has been an active field of research since the 1990s (Tomppo 1991, McRoberts et al. 2002, Reese et al. 2002, Tomppo et al. 2008). The multi-source NFI of Finland is one example where forest area by the dominating tree species groups spruce, pine and deciduous species was among the mapped variables using Landsat images (Tomppo et al. 2009). Gjertsen (2007) used Norwegian NFI data, a Landsat image and a site index map for mapping dominant species groups and age classes in southeastern Norway. NFI-based forest maps can be used in various ways, usually after aggregation to some larger level such as municipalities. However, purely map-based (synthetic) estimates can be severely biased (e.g. Gjertsen 2007), if mapping errors cannot be corrected (Tomppo et al. 2009). Poststratification (Katila et al. 2000, Haakana et al. 2020) or model-assisted estimation (McRoberts et al. 2010) may in such cases be better alternatives that also provide estimates of uncertainty.

The aims of this study were to i) map forest area by the dominating tree species groups spruce, pine, and deciduous at national level for Norway using multi-temporal Sentinel-2 images and other auxiliary data on a fine scale, and ii) to use the map for improving the precision of forest area estimates by tree species for all of Norway and for smaller sub-populations (municipalities). Secondary objectives were to analyze the map accuracy given varying forest structures and to separate the uncertainty in estimates induced by errors in forest maps from errors in tree species maps. Furthermore, the applicability of model-assisted estimation and poststratification were compared. Empirical models were fit using Norwegian National Forest Inventory (NFI) data as the reference. The resulting map was validated using an independent data set consisting of more than 45,000 forest stands where dominant species were determined using aerial stereo images interpretation by experts.

## 3   Material and methods

### 3.1   Study area

The study area is Norway, excluding the non-forested arctic island groups Svalbard and Jan Mayen. The geographical distribution of the main species groups is shown in Figure 1. While the species of economic interest, Norway spruce (*Picea abies* (L.) Karst.) and Scots pine (*Pinus sylvestris* L.), make up for the majority of biomass and volume, the most abundant tree species is downy birch (*Betula pubescens*, Ehrh.) (Table 1). Downy birch also constitutes the main tree species towards most of the forest and tree line in the mountains and the arctic. Although Norway spruce, Scots pine, and downy birch represent 90% of the timber volume, there can be found at least 30 different tree species in Norwegian forests (Table 1).

The forest line is around 1100 m a.s.l. in southern Norway and decreases to 0 m a.s.l. in the very northern tip of Norway. Birch has the largest geographical spread compared to the other species and is found from 58° northern latitude in southern Norway to beyond 70° in the north and from 5° longitude in western Norway to beyond 30° in the north-east. All broadleaved tree species will be considered as one group in this study, as is common in forest inventories in the Nordic countries (Tomter et al. 2010).

Scots pine occurs up to 70° north and is considered as the pine group together with some very rare occurrences of lodgepole pine (*Pinus contorta*). Besides its economic interest in southern Norway, pine forest can be of high ecological value (Øyen et al. 2006). In this respect, it is worth mentioning a large Scots pine population in and around the Pasvik national park at the border to Russia and Finland around 30° east.

Norway spruce has its natural distribution in south-eastern and mid Norway up to approximately 67° north. However, due to its high economic value, spruce was also planted along the west-coast up to almost 70° north (Figure 1), and less commonly also beyond. Along Norway's west-coast, sitka spruce *(Picea sitchensis)* plantations were established, mostly in the 1950s and 1960s. Sitka spruce and all other coniferous tree species, except for pine species, are considered as the tree species group spruce.

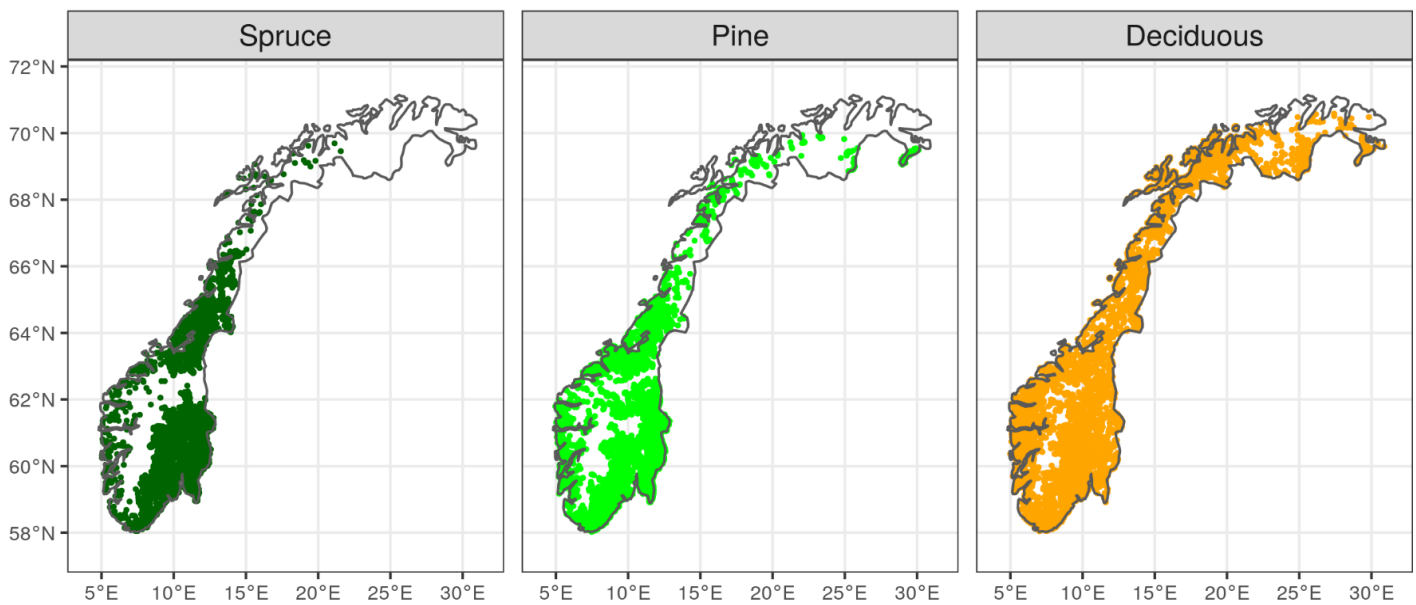

*Figure 1: Geographical distribution of the three main tree species groups spruce, pine and deciduous in Norway. The points are approximate locations of NFI sample plots within forest.*

Table 1: List of the most common tree species in Norway by their timber volume proportion, based on NFI data.

| Species | Biomass proportion (%) |
|---|---|
| Picea abies | 42.3 |
| Pinus sylvestris | 30.1 |
| Betula pubescens | 17.3 |
| Alnus incana | 1.9 |
| Populus tremula | 1.8 |
| Salix caprea | 1.1 |
| Sorbus aucuparia | 1.0 |
| Betula pendula | 1.0 |
| Picea sitchensis | 0.9 |
| Other conifers | 0.3 |
| Other broadleaved species | 2.3 |

## 3.2   NFI Data

We used the 22,008 permanent sample plots of the Norwegian national forest inventory (NFI) for training models and for estimating forest area by dominant tree species. The NFI is a continuous inventory system and uses a systematic, inter-penetrating panel design where 1/5$^{th}$ of the plots are measured every year (Tomter et al. 2010). Stratification is used to sample more observations on a denser sampling grid in potentially productive areas where conifers can be of economic interest, than in low-productive areas. Thus, four design strata are utilized in the NFI; two in the northern-most county of Finnmark (approximately east of 22° eastern longitude), and two outside Finnmark that cover most of Norway's area (Figure 2). For simplicity, we will refer to the latter as being located south of Finnmark. The sampling grid in the area south of Finnmark has a size of 3 km × 3 km in the lowland region (stratum 1) and 3 km × 9 km in the low-productive mountainous region (stratum 2). In the latter, coniferous trees and active forest management seldomly occur (Table 2). Stratum 1 covers approximately 46% of Norway's land area and more than 80% of the country's forest. Therefore, most of the NFI plots within forest are located within stratum 1. In Finnmark, a 3 km × 3 km grid is used within regions where pines can form dense stands of economic interest (stratum 3) and a 9 km × 9 km grid is used outside these regions (stratum 4). As opposed to the rest of the country, the coniferous region is not determined by elevation but by local topography. Norway spruce occurs seldomly in Finnmark. While spruce, pine, and deciduous forests cover approximately equal parts in stratum 1, almost 95% of the forests in stratum 4 are dominated by deciduous trees. The forest proportion in the different strata ranges from 9.5% to 80.0% of which up to 0.8% are unstocked areas. While the two coniferous tree species stand for the majority of biomass in Norway (Table 1), deciduous trees (especially birch) dominate the majority (42%) of the forest area (Table 2). Because the sampling fraction differs among the strata, the plots represent different portions of Norway's land area (sampling weights). The sampling weights for plots in strata 1-4 are 9 km$^2$, 27 km$^2$, 9 km$^2$, and 45 km$^2$.

While the Finnmark strata 3 and 4 are based on a map of productive and non-productive forest, the decision on whether a plot belongs to stratum 1 or 2 was taken differently. For the municipalities along the west-coast and northern Norway (north of the Saltfjell mountain range), areas above a certain elevation threshold were defined as mountainous. These thresholds range from 100 m to 800 m and were defined in cooperation with the local forest administration. Many municipalities in southern and south-eastern Norway do not contain high mountains. For the remaining area (approximately 40% of the country), the stratum allocation was made by the field staff that visited each sample plot on a 3 km × 3 km grid. If a field plot showed characteristics of a mountain forest, this was recorded, and the plot was only included in the sample, if it aligned with the 3 km × 9 km grid. For this area, a stratum map based on universal Kriging using logit-transformed elevation as the covariable was created. A spherical variogram model used to this end had a zero nugget, a range of 5600 m and a sill of 0.73. The model was used to interpolate among the full 3 km × 3 km grid with stratum information at each plot to create a 16 m x 16 m resolution map aligning with the forest resource map SR16 (Astrup et al. 2019).

*Table 2: Forest area by dominating tree species group in the four NFI strata and all of Norway.*

| Stratum | Spruce (%) | Pine (%) | Deciduous (%) | Unstocked (%) | Area (km$^2$) | Forest (%) | # plots in forest | Forest area (km$^2$) |
|---|---|---|---|---|---|---|---|---|
| 1 | 34.0 | 33.7 | 31.8 | 0.6 | 149,885 | 66.4 | 11,046 | 99,551 |
| 2 | 3.6 | 8.1 | 87.8 | 0.5 | 125,281 | 9.5 | 442 | 11,906 |
| 3 | 0.0 | 64.2 | 35.8 | 0.0 | 1,350 | 80.0 | 120 | 1,080 |
| 4 | 0.0 | 4.2 | 94.9 | 0.8 | 47,266 | 20.5 | 118 | 9,683 |
| Norway | 28.0 | 29.1 | 42.3 | 0.6 | 323,782 | 37.7 | 11,726 | 122,220 |

The permanent sample plots of the NFI are circular with a size of 250 m$^2$ and their center coordinates are measured using a handheld GPS. For approximately 74% of the sample plots in forest, the center coordinates have been measured using survey-grade GPS equipment in an effort to measure all center coordinates within the next years. All trees with a diameter at breast height (dbh) ≥ 5 cm are recorded for dbh and tree species. Tree heights are measured for a subset of approximately 10 trees per plot and the heights of the remaining trees are estimated utilizing the measured heights. Species-specific volume equations are used to estimate the volume of each recorded tree. For each tree-species group, timber volume was totaled on the plot-level and scaled to per ha values. Plot-level volume proportions were used to calculate the dominant tree species, which is the variable of interest in this study.

For each sample plot, the dominant tree species of the stand is also assessed (determined visually) by the field staff on a circular 0.1 ha plot concentric to the 250 m$^2$ sample plot. We use this information for plots that were located in forest but did not have tree recordings (Table 2). This is often the case in young forest where trees have not reached the dbh threshold or, less commonly, in sparse forests without trees on the 250 m$^2$ plot.

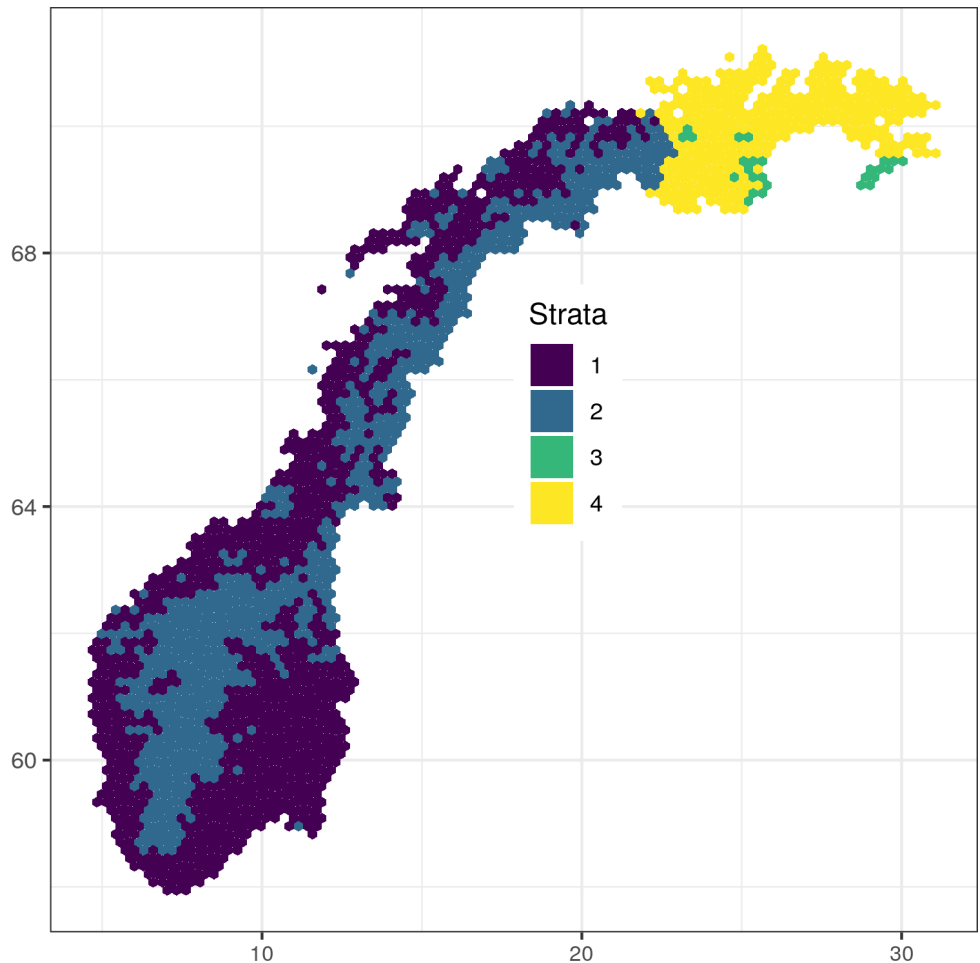

*Figure 2: NFI stratum map (aggregated to large pixels for display).*

### 3.2.1 NFI data used for modelling and estimation

Data from the 2014-2018 NFI cycle were used and the dominant tree species group was determined based on which group had the most timber volume at a plot. Of the 22,008 sample plots, 11,726 had a center in forest. Of those, 11,661 plots were within stocked forest and dominant species was identifiable either by plot-level timber volume proportions or by the stand-level assessment by the field staff. These plots were used for estimating (the stocked) forest area by dominant tree species (Section 3.7). The other 65 plot centers were mostly located on temporarily unstocked areas, typically shortly after harvests without regeneration. Of the plots in stocked forest, 10,544 were unsplit; that is, the full plot radius was located within one forest stand. The other plots were either split between two forest stands (477 plots) or between forest and other land uses (640 plots). Of the unsplit plots, dominant tree species was identified by the plot-level timber volume proportions for 10,279 plots. These sample plots were used in the modelling process (Section 3.6). For the remaining 265 sample plots, dominant tree species was identified by the stand-level estimate in the field. These plots were typically young forests where the regeneration did not yet reach the 5 cm DBH threshold or sparse low-productive forests. All 22,008 sample plots were used for estimating forest area by tree species.

## 3.3 Independent validation data

Reference data for an independent validation were stand-level observations from 14 forest management inventories (FMIs) conducted in 2003 (one FMI) and between 2012 and 2018. The FMIs were located between 58 and 65 degrees North and cover the area that is most relevant for forest management in Norway (Figure 3). FMIs in Norway are conducted by specialized inventory companies under assignment of a forest owner or forest owner's association. The first step in an FMI is the manual delineation of stands using aerial stereo images and the determination of tree species group proportions (spruce, pine, and deciduous) within each stand. We selected pure stands, where one tree species covered more than 95% of the stand area to avoid uncertain reference data. This data set of pure stands comprised of 42,966 stands covering an area of more than 756 km$^2$.

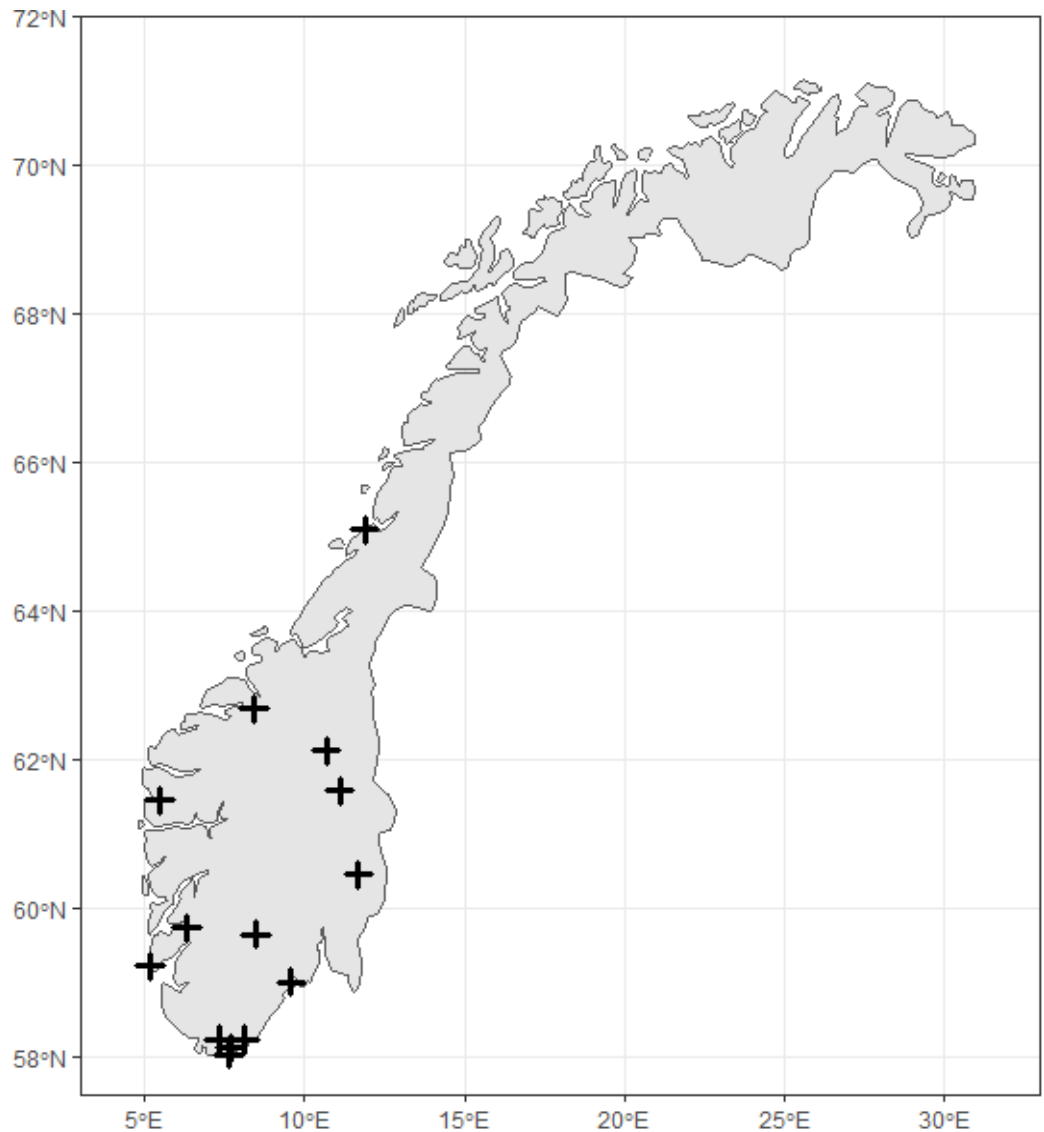

*Figure 3: Location of forest management plans (crosses) within Norway.*

## 3.4 Auxiliary data

### 3.4.1 Remotely sensed data

Two cloud- and shadow-free Sentinel-2 bottom of atmosphere (L2A) mosaics were obtained from the SentinelHub (Kirches 2018). One mosaic was based on images acquired in May 2018 (leaf-off conditions), the other mosaic was based on images acquired in July 2018. The medoid method (Flood

2013) was used to obtain the most representative ground surface pixel value within each month, thus avoiding clouds and haze. The mosaics contained the bands B1, B2, B3, B4, B5, B6, B7, B8, B8A, B11, and B12 and had a 10 m pixel size. Bands with original resolution coarser than 10 m (all bands except B2-B4) were up-sampled to 10 m using the nearest neighbor method. As an additional predictor variable, the NDVI was computed for each mosaic.

Predictor variables for modelling were calculated by overlaying the mosaic with the NFI plots and extracting the area-weighted means of the pixels intersecting with the sample plot polygons. For mapping, that is for application of the models, the mosaics were resampled, using bilinear interpolation, to a raster with 16 m × 16 m pixel size that aligns with the Norwegian forest resource map SR16 (Astrup et al. 2019).

For a total of nine and three of the NFI plots within forest, the May and July mosaics within forest were empty. To obtain predictor variables for all NFI plots, the radius for calculating predictor variables was extended to 18 m. For mapping, empty pixels were ignored and were thus implicitly mapped as non-forest.

### 3.5 Data from existing maps

The Norwegian National Land Resource Map, AR5 (Ahlstrøm et al. 2019), was developed in the 1960s by intensive field work at a scale of 1:5,000. AR5 contains, among others, information on forest/non-forest and forest type as the classes coniferous dominated, deciduous dominated and mixed forest. AR5 is available for most of the productive forest. For the remaining land area, AR5 was complemented with forest/non-forest information from N50, Norway's official topographical map at a scale 1:50,000 (Kartverket 2018). Forest type in the regions that were complemented with N50 was set to the class deciduous dominated forest. The combined AR5 and N50 vector map was converted to a 16 m × 16 m raster that aligns with the Norwegian forest resource map SR16 (Astrup et al. 2019). We will refer to the combined AR5 and N50 raster as *AR5 forest type*. AR5 forest type was used as a predictor variable in the models and in mapping by applying the models only for grid cells with AR5 forest-type information (i.e., within forest). That means, AR5 forest type was used as a forest mask. The overall accuracy of the AR5 forest mask was 92% (Table 3, Table 4).

*Table 3: Confusion matrix of the AR5 forest mask (all values in %).*

|  |  | NFI |  |  |
|---|---|---|---|---|
|  |  | Non-forest | Forest | User's accuracy |
| AR5 | Non-forest | 58 | 4 | 94 |
|  | Forest | 4 | 34 | 89 |
|  | Producer's accuracy | 93 | 89 |  |
|  | Overall accuracy |  |  | 92 |

Table 4: Accuracy of the AR5 forest mask in the four NFI strata (all values in %).

| Stratum | Overall accuracy | Producer's accuracy non-forest | Producer's accuracy forest | User's accuracy non-forest | User's accuracy forest |
|---------|------------------|-------------------------------|----------------------------|----------------------------|------------------------|
| 1 | 89 | 85 | 91 | 83 | 92 |
| 2 | 95 | 98 | 72 | 97 | 77 |
| 3 | 93 | 87 | 95 | 81 | 97 |
| 4 | 90 | 90 | 91 | 97 | 69 |

Information on elevation (height above sea level) was obtained from the Norwegian DTM10 (Kartverket 2018), a digital terrain model based on aerial stereo images interpretation with a vertical accuracy of 2-6 m and a 10 m grid cell size. Finally, the DTM10 was resampled using bilinear interpolation to a 16 m × 16 m raster.

## 3.6 Modelling strategy

Random Forest models (Breiman 2001) were "fit" by adding and removing predictor variables in a forward and backward selection scheme. The response variable was the dominant tree species (spruce, pine, or deciduous) observed on 10,279 NFI sample plots fully in forest (Section 3.2.1). First, variables were added if they improved the overall accuracy (OA). Once all variables were tested, variables were removed iteratively, starting with those that had the lowest variable importance. If the removal of a predictor variable did not reduce the OA, the variable was finally removed from the model.

Three types of models were fit consisting of different sets of potential predictor variables. I) A model based on Sentinel-2 variables obtained from the July 2018 mosaic, elevation, and approximate longitude and latitude. II) A model based on the variables in group I) with additional Sentinel-2 variables obtained from the May 2018 mosaic. III) A model based on the variables in group II) with an additional variable on coniferous, deciduous, or mixed stands from the AR5 map. For simplicity, we will refer to the models with variables selected from the three sets as models I, II, and III.

In an initial analysis, the two RF parameters ntrees (number of trees) and mtry (number of predictor variables to consider in each node) were tuned. Model III with variables selected using the default parameters (ntrees=500 and mtry=p/3 with p as the number of predictor variables) was utilized to this end. The parameter ntrees was varied between 100 and 1000. A slight increase of the OA was visible between 100 and 400 trees, but not beyond the default of 500, which was therefore kept for all analysis. The parameter mtry was varied between p/1 and p/10. A slight increase of the OA was visible between p/1 and the default p/3, which was therefore kept for all analysis.

The sample plots in the four different NFI strata are laid out along grids with different sizes (see Section 3.2) and therefore represent different areas of land which is reflected in different sampling weights. Therefore, overall accuracy (OA), producer's (PA), and user's (UA) accuracy were computed by considering the sampling weights. The model with the greatest overall accuracy, based on all sample plots with a center in forest, was further analyzed and used for the estimation of forest area by dominating tree species group. For simplicity, the latter will be denoted spruce, pine, or deciduous forest.

## 3.7 Estimation of forest area by dominating tree species group

A direct (expansion) estimate based on NFI data for the forest area or forest area dominated by a tree species is

$$\hat{t} = \sum_{i \in s} y_i / \pi_i = \sum_{h=1}^{H} \hat{t}_h = \sum_{h=1}^{H} \left( A_h \frac{1}{n_h} \sum_{j \in s_h} y_j \right), \quad i = 1, \ldots, n, j = 1, \ldots, n_{s_h} \quad (1)$$

where $s$ is the sample, $\pi_i$ is the inclusion probability (inverse of the sampling weight) of the plot that, in our case, differs among the strata but is constant within a stratum, and $y_i$ is an indicator variable for sample plot $i$ that is 1 if the sample plot center is in domain $d$ of interest and 0 otherwise with d = {forest, spruce forest, pine forest, deciduous forest}. For example, when estimating the forest area dominated by spruce, d will be one for all sample plots with a center in spruce forest and zero for all other plots. As can be seen from the second and third terms, this estimator can also be given as a stratified estimator where h is the stratum index, $H=4$ is the total number of strata, $A_h$ is the stratum area, and $n_h$ is the number of sample plots in a stratum.

The stratified design of the NFI has to be considered to estimate the variance

$$\hat{V}(\hat{t}) = \sum_{h=1}^{H} \hat{V}(\hat{t}_h) \quad (2)$$

where $\hat{V}(\hat{t}_h)$ is the variance within a stratum

$$\hat{V}(\hat{t}_h) = A_h^2 S_h^2 / n_h \quad (3)$$

with $S_h^2$ as the sample variance

$$S_h^2 = 1/(n_h - 1) \sum_{i \in s_h} (y_i - \bar{y}_h)^2 \quad (4)$$

and $\bar{y}_h$ is the mean of the observed indicator variables. The standard error is the square root of the variance

$$SE(\cdot) = \sqrt{\hat{V}(\cdot)}. \quad (5)$$

Model-assisted estimation and poststratification (Särndal et al. 1992, McRoberts 2011) are used to include the mapped tree species information in the estimation process and will result in reduced variance estimates, if the map represents the tree species observed at the sample plots well. The model-assisted estimator is given by

$$\hat{t}_{MA} = \tilde{t} + C \quad (6)$$

where $\tilde{t}$ is the map-based (synthetic) estimate of the *d*-th domain area obtained by adding the pixel areas of the *d*-th domain and *C* is a correction term similar to (1), given by

$$C = \sum_{i \in s} e_i/\pi_i \qquad (7)$$

with

$$e_i = y_i - \hat{\hat{y}}_i \qquad (8)$$

the residual where $\hat{\hat{y}}_i$ is the cross-validated prediction within the forest mask. This means, if the forest mask and the species prediction for plot i are correct, the residual will be 0. If either the forest mask or the species prediction for plot i are wrong (different to the NFI observation), the residual will not be 0 and contribute to an increased variance.

The variance of the model-assisted estimate $\hat{V}(\hat{t}_{MA})$ is given by eq. (2) after substituting the sample variance of the observations with the sample variance of the residuals

$$S_h^2 = 1/(n_h - 1) \sum_{i \in s_h} (e_i - \bar{e}_h)^2 \qquad (9)$$

where $\bar{e}_h$ is the mean residual.

Poststratification (PS) can be more precise than MA estimation, if the auxiliary information is categorical as is the case for a tree species map (McRoberts et al. 2016). The stratified estimator can be used for PS (Thompson 2002, p. 124) within each NFI design stratum h by

$$\hat{t}_{PS} = \sum_{h=1}^{H} \hat{t}_{PS,h} \qquad (10)$$

and

$$\hat{t}_{PS,h} = \sum_{g_h=1}^{G_h} A_{g_h} \bar{y}_{s_{g_h}} \qquad (11)$$

where PS indicates poststratification, $A_g$ is the mapped area of group g within stratum h and $\bar{y}$ is the mean of the indicator variables $y_i$ observed on plots within group g. The groups are given as a binary map where one group is the domain d of interest and the other group comprises of all other domains including non-forest. The variance of this estimate results from using eq. (2) with

$$\hat{V}(\hat{t}_h) = \hat{V}(\hat{t}_{PS,h}) = \sum_{g_h=1}^{G_h} A_{g_h}^2 S_{g_h}^2 / n_{g_h}. \qquad (12)$$

## 3.8 Application of the estimators

The estimators described in the previous section can also be applied to any smaller sub-population such as a region or municipality that contains a sufficient number of sample plots. Due to the foundation on the central limit theorem, a minimum of 30 observations in MA is sometimes mentioned (Thompson 2002, p. 76) and with 20 observations per group "one should be on the safe side" in the case of PS (Särndal et al. 1992, p. 267). That means, PS based on two groups would require at least 40 observations if the two groups were exactly balanced (i.e. each group covers 50% of the area). Consequently, the required minimum number of observations is a fair bit higher for PS than for MA because the groups are seldomly balanced in practice. In our case, the minimum number of plots is required per NFI stratum and the number of observations includes plots in forest and non-forest.

The relative efficiency (RE) was used to quantify the improvement of a MA or PS estimate compared to a direct estimate. The RE is the ratio of the variances $RE = \hat{V}(\hat{t})/\hat{V}(\hat{t}_{\{MA,PS\}})$. Assuming simple random sampling, its interpretation is a factor for the number of sample plots to achieve the same variance using the direct estimator as with the MA or PS estimator. For example, a RE of 1.5 means, if the number of sample plots was increased by 50%, the direct estimator would result in the same variance as the MA or PS estimator.

As opposed to the MA area estimate, the synthetic (map-based) estimate is not needed to estimate the variance of the MA area estimate. Therefore, it was possible to separate the variance contribution of the forest mask and the tree species map by estimating the MA variance (eq. (9)) with the residual

$$e_i = y_i - \hat{y}_i \tag{13}$$

where $\hat{y}_i$ is the 10-fold cross-validated prediction for plots that are forest according to the NFI definition. This results in variance estimates that could be obtained if an exact forest mask was available.

If the species model and the forest mask were error-free, all residuals $e_i$ would be 0, which means that the variance would be 0 and $\hat{t}_{MA} = \tilde{t}$. The latter would also be the case, if the prediction errors were exactly balanced (as many false positives as false negatives), the variance, however, would not be 0 in this case. If the species map or the forest mask had systematic errors, this would be corrected for by the correction term $C$.

The area estimate of spruce, pine, and deciduous forest, and forest area (eq. (6)) will be given for the whole country. For smaller areas, such as strata or municipalities, we only report the variances or RE of MA or PS estimates. This is because our main interest is to see, if estimates improve, when using maps in addition to NFI sample plots. Area estimates are of course of main interest for management and decision support.

It may be noted that the applied estimators overestimate the variance because they assume simple random sampling but are applied to a systematic NFI sample (e.g. Magnussen and Nord-Larsen 2019). While alternative estimators are available, we will not consider them here. Nonetheless, the positive effect of using auxiliary information will also be visible when using these alternative estimators (Magnussen et al. 2018). Tree species areas can be presented as proportions of the forest area. The ratio estimator utilized to this end (Mandallaz 2007, p. 64), however, does not profit from more precise area estimates. We therefore do not focus on forest area proportions.

# 4 Results

## 4.1 Variable selection

The overall accuracy (OA), was 77.5%, 78.6%, and 79.4% for model I (selection among Sentinel-2 variables from July 2018), model II (model I variables and Sentinel-2 variables from May 2018), and model III (model II variables and AR5 forest type), respectively. Similarly, the producer's (PA), and user's (UA) accuracies increased when including Sentinel-2 variables from May to the model that included variables from July only (Figure 4). Except for a slight reduction in the UA of spruce, the same was observed when including the AR5 forest type. All accuracies given are based on a 10-fold cross-validation of the 10,279 sample plots used in the models. The differences between cross-validated and non-cross-validated OA were less than 1%.

The number of selected predictor variables ranged between 11 and 12. The most important variable in all models was Sentinel-2 band B6 from July (Figure 5). Also, bands B4 and B8A from July were among the most important variables in all models. The spatial variables approximate latitude, approximate longitude, and elevation were selected in all models and were of average importance. Their removal (results not shown) would have resulted in considerable amounts of commission errors for spruce in Finnmark (northern Norway) where spruce is not a stand-dominating species. Sentinel-2 band B8A and NDVI from May were the only additional variables in model II (Figure 5) and replaced Sentinel-2 band B1 from July. The variable AR5 forest type was the second-most important variable in model III and replaced Sentinel-2 band B2 from July (Figure 5).

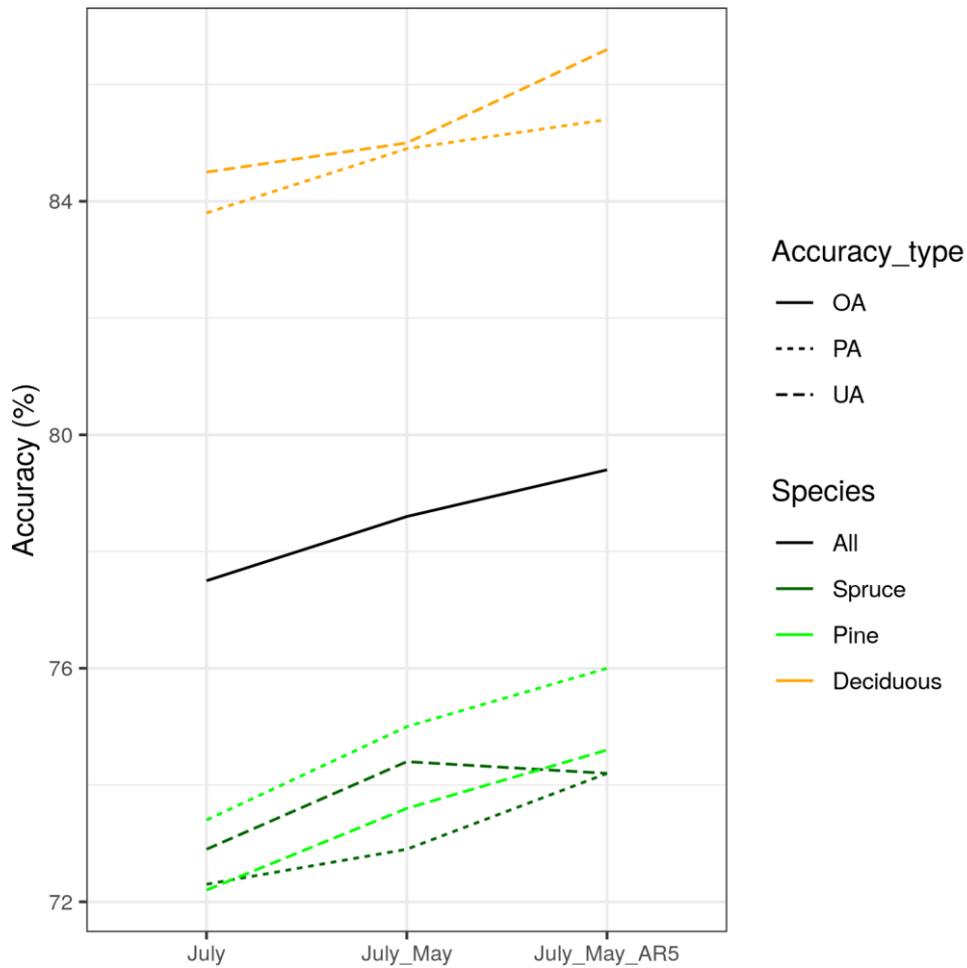

Figure 4: Overall (OA), producer's (PA), and user's (UA) accuracies for models based on Sentinel-2 variables from July (model I); July and May (model II); and July, May and the AR5 forest type map (model III).

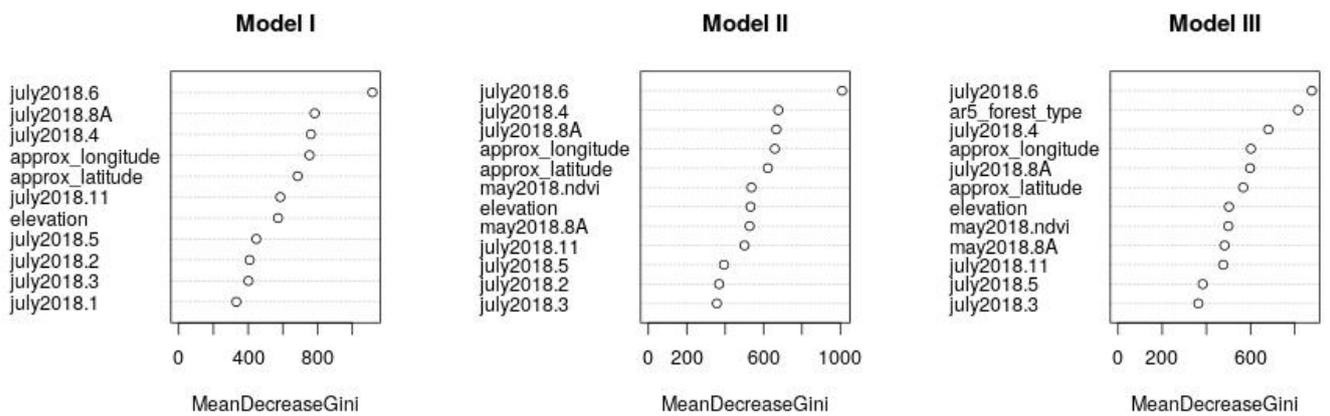

Figure 5: Predictor variables (y-axis, e.g. July2018.6 = B6 of Sentinel-2 mosaic from July) and their importance (Mean decrease in Gini).

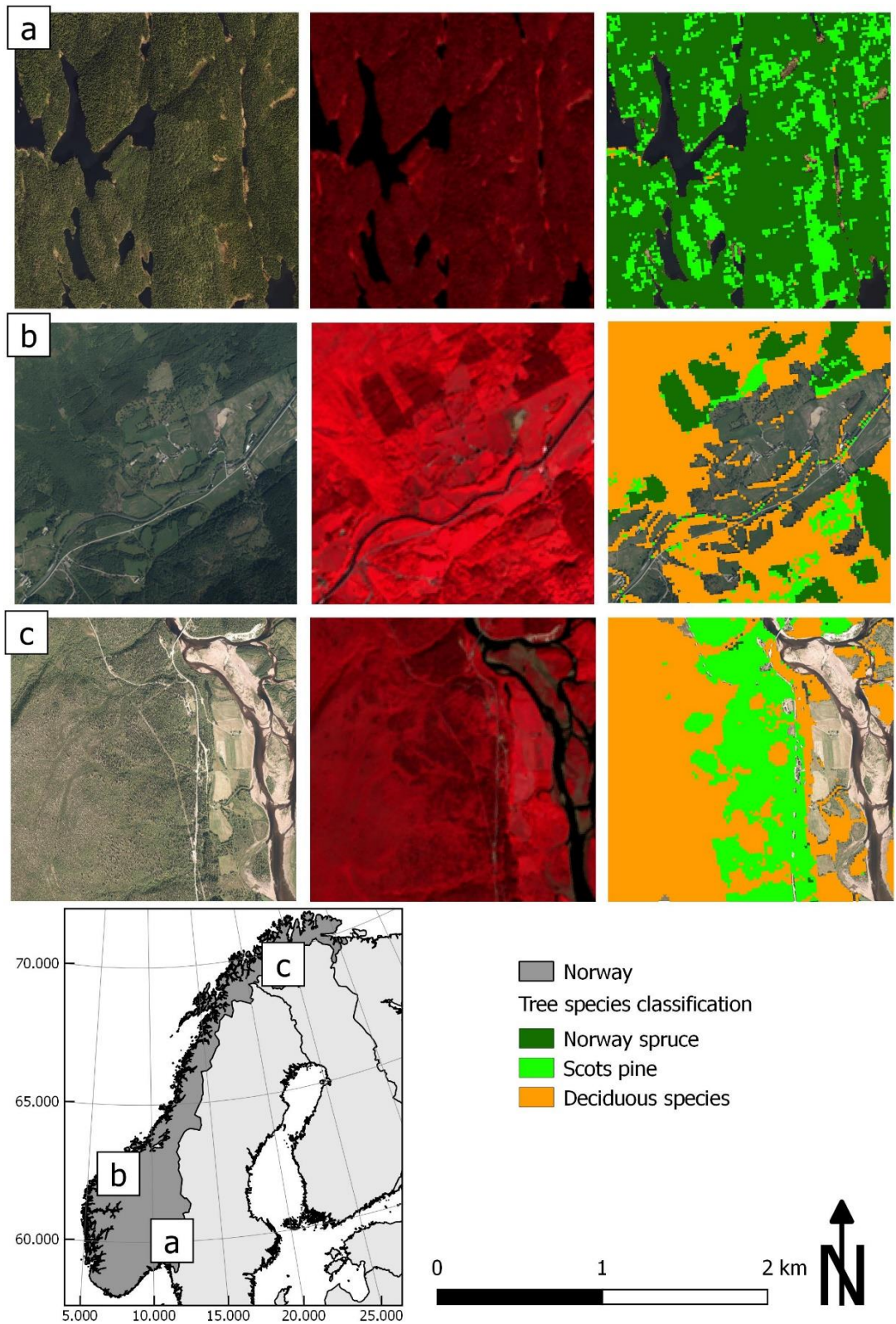

*Figure 6: Orthophoto (left column), Sentinel-2 image (mid column, red= B8; green= B4; blue = B3), and prediction map based on model III (right column) from three subsets (a-c) in Norway.*

## 4.2 Model accuracy

Model III, i.e., the model including Sentinel-2 variables from July and May, and AR5 forest type, had the greatest OA and was used for mapping and subsequent inference on forest areas dominated by a tree species group. Visually, a map based on model III often correspond well with orthophotos. An example of the map from the spruce and pine dominated area in south-eastern Norway is in Figure 6, a. An example from the west coast shows a clear distinction of spruce plantations from the surrounding deciduous forest (Figure 6, b). The distinction between productive pine forest and low-productive birch forest can be seen in Figure 6, c. The accuracy of model III under different conditions was analyzed further. From its confusion matrix (Table 5) it can be seen that misclassifications were more likely to occur between the two coniferous species pine and spruce than between them and the deciduous class. In general, however, the omission and commission errors among the classes were rather balanced.

*Table 5: Confusion matrix of model III (all values in %).*

|  |  | Reference |  |  |  |
|---|---|---|---|---|---|
|  |  | Spruce | Pine | Deciduous | User's accuracy |
| Prediction | Spruce | 20.8 | 4.3 | 3.0 | 74.2 |
|  | Pine | 4.7 | 22.9 | 3.1 | 74.6 |
|  | Deciduous | 2.6 | 3.0 | 35.7 | 86.6 |
|  | Producer's accuracy | 74.2 | 76.0 | 85.4 |  |
|  | Overall accuracy |  |  |  | 79.4 |

The OA of the model in the different NFI strata ranged between 76% and 97% (Figure 7). In stratum 1, the largest stratum that covers the most productive lowland forests (Figure 2), all tree species occurred by approximately equal parts (Table 2). The accuracies in this stratum varied less, ranging from 74% to 80%, compared to the other strata. The slightly higher PA than UA for pine and the opposite for deciduous forest indicates that the model had a tendency to favor pine over deciduous forest in this stratum. The smallest PA (29%) was observed for spruce in stratum 2 (the mountain forest stratum), because most of the rare spruce forests were predicted as deciduous forests by the model. The relatively high OA of 91% in this stratum can be explained by the high prevalence of deciduous forests that were mapped quite well (PA=93%, UA=96%, Figure 7). Spruce did not dominate any NFI plots in in the two Finnmark strata in northern Norway (strata 3 and 4) and does therefore not occur in these strata in Figure 7. In stratum 3, which covers the relatively small coniferous region in Finnmark, PA and UA of pine and deciduous had opposite tendencies because the model tended to favor deciduous over pine forests. This was a general tendency of the model in strata 2, 3, and 4: The UA of deciduous forest was smaller than the PA because the model tended to favor deciduous forest.

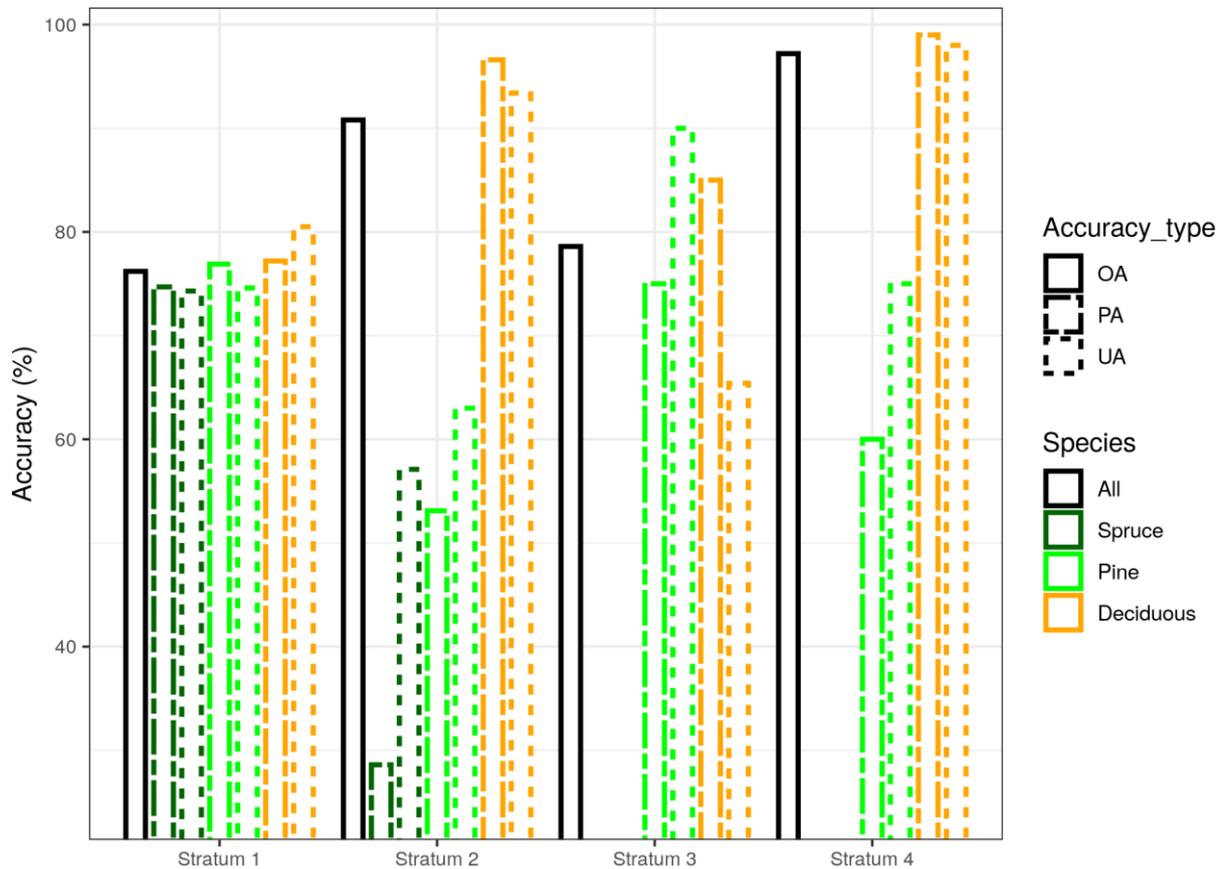

*Figure 7: Overall (OA), producer's (PA), and user's (UA) accuracies of model III by NFI strata (Stratum 1: lowland forest; Stratum 2: mountain forest; Stratum 3: productive forest in Finnmark; Stratum 4: low-productive forest in Finnmark).*

The OA of model III increased from 64% to more than 80% with increasing stand age-class (Figure 8). Because age-classes are not meaningful for unmanaged forest and not all tree species occurred in all strata, this analysis was limited to stratum 1 (i.e., lowland forest south of Finnmark). While the PA and UA increased for pine and deciduous forest from young forest to old forest, the accuracies of spruce levelled out and even decreased for old forest. For forest in the regeneration phase, the model had a tendency to favor pine over deciduous forest which resulted in small UA and PA for pine and deciduous forest, respectively. In the older age-classes, omission and commission errors more likely occurred between spruce and pine. For old forest, the model had a slight tendency to favor pine over spruce. The opposite tendency occurred for young forest and production forest (Figure 8).

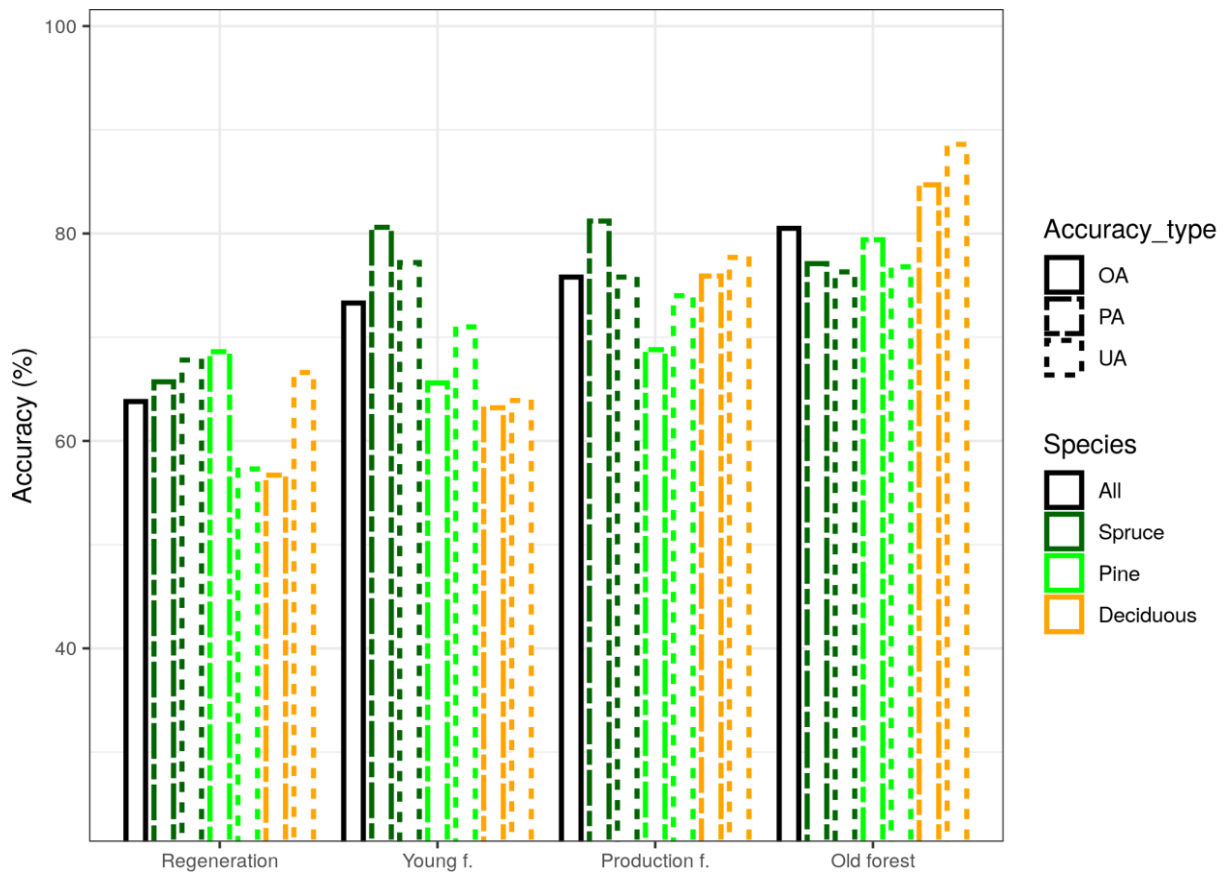

*Figure 8: Overall (OA), producer's (PA), and user's (UA) accuracies of model III) in stratum 1 by age class.*

For estimating forest area by tree species, predictions are also needed for the 1,447 sample plots that have a center in forest but were not included in the model (12.6% of the forested plots). These were plots with no trees measured on the sample plot but a stand-level species assessment, completely unstocked stands shortly after harvests, and split plots. The relatively high OA for plots with a stand-level species assessment resulted from the high prevalence of deciduous forests, which is to be expected for forest that did not yet reach the dbh threshold (Figure 9). Of the 65 completely unstocked plots, 19%, 36%, and 44% were mapped as spruce, pine, and deciduous forests, respectively. The split plots represent pixels that cover two different stands or forest and non-forest. The accuracies for these plots were, as can be expected for mixed pixels, much smaller than for the plots within homogenous forest used for modelling and ranged between 40% and 86% (Figure 9). In both cases, the model tended to favor pine over the other species. This tendency was especially strong for spruce forests close at the border to non-forest. This may in part also explain the observed tendency to favor pine in old forests (Figure 8) because these forests often are open with many canopy gaps which can resemble internal stand borders. The PA and UA using all 22,008 NFI sample plots (10,279 plots used for modelling, the 1,447 plots considered in this paragraph, and 10,282 plots in non-forest) assuming an exact map for forest and non-forest areas were available, ranged between 71% and 86% (Table 10).

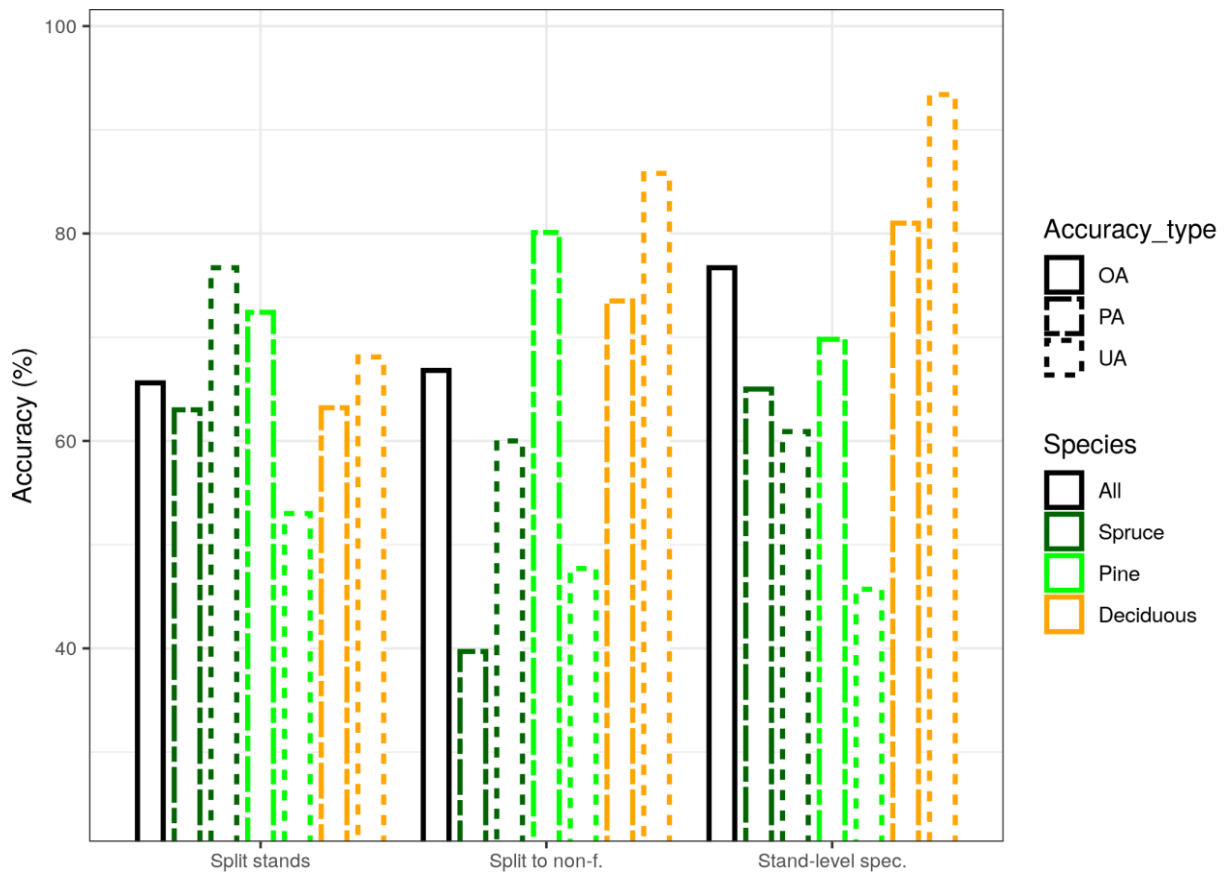

*Figure 9: Overall (OA), producer's (PA), and user's (UA) accuracies of model III predictions by plots split between two stands, split between forest and non-forest, and for plots where dominant species was based on the stand-level assessment.*

For national mapping and estimation, in addition to a tree species map, a forest (boundary) map (i.e. a forest mask) is needed that introduces additional uncertainty, unless the map is without error. We used the AR5 forest mask which was relatively precise but not error-free (Table 4). The OA based on all 22,008 NFI including sample plots not used in the model and the non-forest class resulting from the AR5 forest mask was 84%. Due to errors in the forest mask and the uncertainty introduced by plots not used in the model (Figure 9), the PA and UA reduced to 65%-71% for the tree species map on the national scale (Table 6). To give an impression of the accuracy on a sub-national scale, accuracy measures were calculated for each of the 10 counties in Norway. As could be expected, the range of the PA and UA within the counties was even wider than on the national level (Figure 10, Table 11). The extremes were the PA and UA of 22% and 100% for spruce forest in northern Norway where spruce hardly occurs. Otherwise, the PA and UA ranged between 44% and 79%.

Table 6: Confusion matrix for model III combined with the forest mask and including NFI sample plots not used for modelling (all values in per-cent (%) of the Norwegian land area as represented by the NFI sample plots).

|  |  | Observed |  |  |  |  |
|---|---|---|---|---|---|---|
|  |  | Spruce | Pine | Deciduous | Non-forest | User's accuracy |
| Predicted | Spruce | 7.4 | 1.5 | 1.1 | 0.4 | 71.1 |
|  | Pine | 1.8 | 7.6 | 1.1 | 1.3 | 64.9 |
|  | Deciduous | 0.8 | 0.9 | 11.3 | 2.9 | 71.3 |
|  | Non-forest | 0.6 | 1.0 | 2.4 | 58 | 93.6 |
|  | Producer's accuracy | 70.2 | 69.3 | 71.1 | 92.8 |  |
|  | Overall accuracy |  |  |  |  | 84.4 |

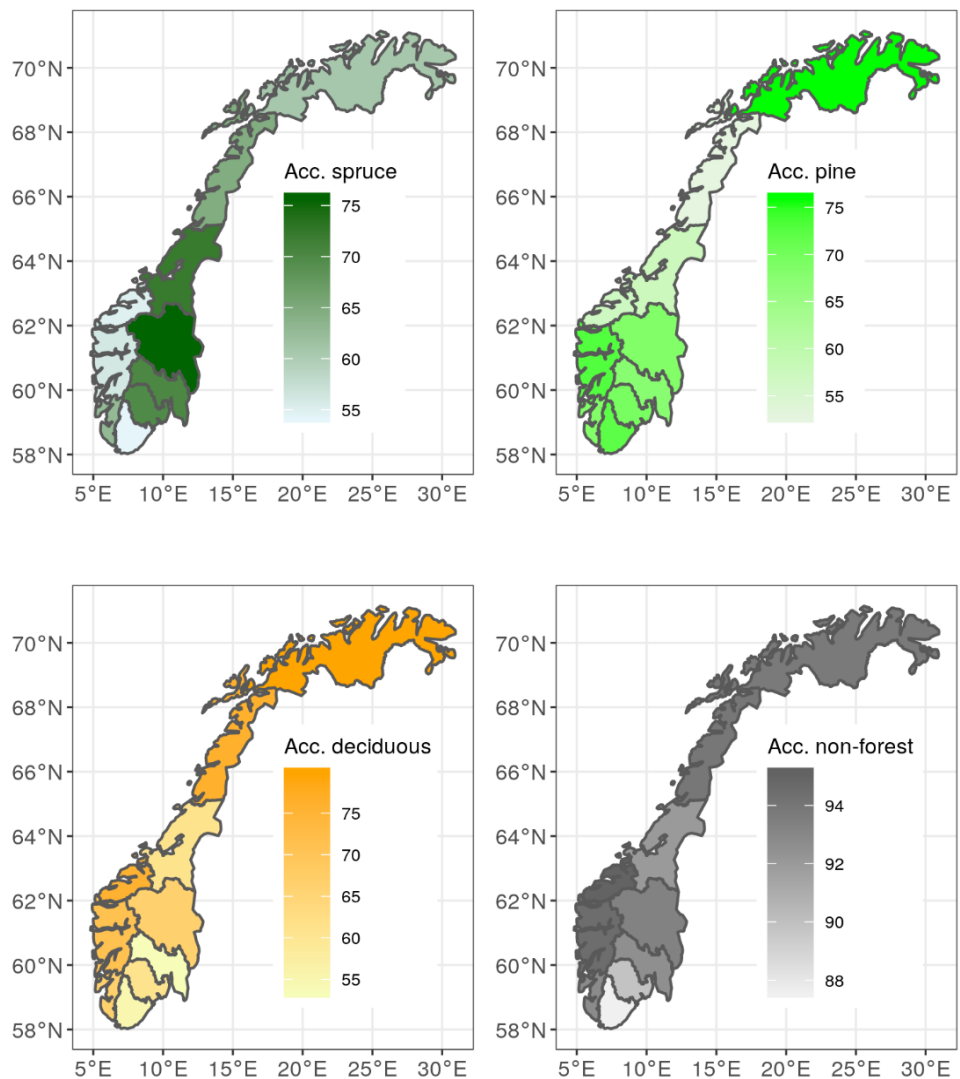

Figure 10: Average of producer's and user's accuracy (Acc.) for spruce, pine, and deciduous forest, and non-forest, in Norway's counties.

## 4.3 Independent validation

The validation stands covered an area of more than 756 km$^2$ in mid- and southern Norway and the overall accuracy of the prediction map based on model III on stand level was 90% (Table 7). As for the plot-level model assessment (Table 5), the mis-classification between spruce and pine stands was greater than between coniferous and deciduous stands (Table 7). However, more than twice as many pine stands were mis-classified as spruce stands than vice versa. The user's and producer's accuracies ranged between 84 and 95% (Table 7). The proportion of deciduous stands in the validation data of 9% was much smaller than in the modelling data set because of little economical interest in deciduous trees which leads to frequent omissions of deciduous stands from forest management inventories.

*Table 7: Confusion matrix of synthetic stand-level estimates based on model III (all values in %).*

|  |  | Reference |  |  |  |
|---|---|---|---|---|---|
|  |  | Spruce | Pine | Deciduous | User's accuracy |
| Estimate | Spruce | 32.1 | 5.5 | 0.3 | 84.7 |
|  | Pine | 2.2 | 50.4 | 0.4 | 95.0 |
|  | Deciduous | 0.4 | 0.7 | 8.0 | 88.3 |
|  | Producer's accuracy | 92.5 | 89.0 | 92.0 |  |
|  | Overall accuracy |  |  |  | 90.5 |

## 4.4 Area estimation

Based on model III, the sum of the synthetic estimates of spruce, pine, and deciduous forest on the national scale was 122 thousand-km$^2$. Due to empty pixels in the Sentinel-2 mosaic, this area was 75 km$^2$ or 0.006% smaller than the synthetic forest area estimate based on the AR5 forest mask. Because of the relatively small difference, no adjustment was applied.

Synthetic estimates of spruce, pine, and deciduous forest were corrected by C={422, -2,379 and 123 km$^2$} to obtain model-assisted (MA) estimates of 35, 34, and 51 thousand-km$^2$, respectively (Table 8). The relatively large correction factor for pine resulted from an overprediction of pine in the lowland stratum that was not levelled out by an underprediction in the other strata. The relative efficiencies (RE) of the MA estimates were 1.3, 1.2, and 1.5 for spruce, pine, and deciduous forest, respectively. This means that between 20% and 50% more sample plots would be needed to obtain the same precision using the direct estimator (based on NFI plots without using the species prediction map). For standard errors, this improvement appeared less impressive (0.1% - 0.4% reduction) because they were already quite small for the direct estimate (1.5% – 2.0%) (Table 8). The RE of MA estimates of coniferous forest (i.e., the combined area of spruce or pine dominated forests) was 1.7, and thus considerably greater than for spruce and pine forests individually, because most of the commission and omission errors occurred between spruce and pine forest (Table 5). The MA estimate of forest area (over all tree species) had the greatest improvement in terms of variance compared to the direct estimate with a RE of 1.8 (Table 8). MA area proportions of forest dominated by spruce, pine, or deciduous species were 29%, 28%, and 42%, respectively. In comparison to the direct estimates, the MA ratios changed by 0.9%, 0.9%, and 0.1%, respectively.

Table 8: Characteristics of direct and model-assisted estimates given AR5 as the forest mask.

|  | Direct estimate $\hat{t}$ (km²) | Coefficient of variation $CV(\hat{t})$ (%) | Correction factor C (km²) | Model assisted estimate $\hat{t}_{MA}$ (km²) | Coefficient of variation $CV(\hat{t}_{MA})$ (%) | Relative Efficiency - RE | Difference $\hat{t} - \hat{t}_{MA}$ (%) |
|---|---|---|---|---|---|---|---|
| Spruce | 34,254 | 1.5 | 423 | 35,346 | 1.3 | 1.3 | -3.9 |
| Pine | 35,626 | 1.5 | -2,378 | 34,020 | 1.4 | 1.2 | 4.5 |
| Deciduous | 51,644 | 2.0 | 123 | 50,785 | 1.6 | 1.5 | 1.7 |
| Forest | 122,222 | 0.9 | -1,138 | 120,923 | 0.7 | 1.8 | 1.1 |

There were neither observations nor map predictions of spruce forest at NFI plots in the Finnmark strata. For MA estimation this means that these NFI plots contribute with zero variance to the national estimate and that the synthetic estimate (i.e. the area of the map predictions) is the MA area estimate. Due to the division by zero, a RE cannot be calculated in the Finnmark strata (Table 9). It is known that spruce was planted in Finnmark and larger patches of mapped spruce forest appeared plausible when visually compared to aerial images. The MA (and synthetic) estimates of spruce forest in Finnmark were 288 ha and 214 ha, corresponding to 0.004% and 0.006% of Finnmarks land area, for stratum 3 and 4, respectively.

Poststratification (PS) was not applicable for spruce forest in the Finnmark strata because of the zero-sized group of predicted spruce forest at NFI plots. Due to the constrain of at least 20 observations per group (see Section 3.8), PS was also not applicable for spruce forest in the mountain stratum. Despite of more than 400 forested plots in the mountain stratum, only 10 had a map prediction for spruce forest, which made this group too small for reliable a PS estimate. The same was true for pine forest in the low-productive stratum in Finnmark where three plots had a map prediction of pine forest. However, for strata that allowed a comparison, the RE of PS estimates were always greater than for MA estimates. The RE of PS in these strata ranged between 1.2 and 2.7 (Table 9). Forest area and deciduous forest had sufficient numbers of sample plots with map predictions in all strata and resulted in REs of 2.2 and 1.9 on the national scale.

Table 9: Relative efficiencies for model-assisted (MA) and poststratified (PS) estimates within strata. (Stratum 1: lowland forest; Stratum 2: mountain forest; Stratum 3: productive forest in Finnmark; Stratum 4: low-productive forest in Finnmark).

| Estimator | MA | | | | PS | | | |
|---|---|---|---|---|---|---|---|---|
| Stratum | 1 | 2 | 3 | 4 | 1 | 2 | 3 | 4 |
| Spruce | 1.3 | 0.9 | 0/0 | 0/0 | 1.6 | - | - | - |
| Pine | 1.1 | 1.0 | 1.3 | 2.5 | 1.5 | 1.2 | 1.6 | - |
| Deciduous | 1.3 | 1.7 | 1.3 | 1.5 | 1.6 | 2.0 | 1.7 | 2.0 |
| Forest | 2.1 | 1.8 | 2.4 | 1.7 | 2.4 | 2.1 | 2.7 | 2.1 |

If an exact forest mask was available, the RE would improve slightly to 1.4 and 1.4 for MA estimates of forest area dominated by spruce and pine (RE of 2.3 for all conifers) and considerably (RE=6.0) for deciduous forest (Table 8). The main reason for this improvement were the relatively large omission and commission errors of approximately 30% of the AR5 forest mask in the strata with low-productive forest (strata 2 and 4, Table 4). The errors of the AR5 forest mask in combination with the small accuracy of model III for spruce in stratum 2 (the mountain stratum) resulted in a RE<1. This means, in this case the direct estimate was more precise than the MA estimate (Figure 11). If an exact forest mask was available, the RE would improve to approximately one.

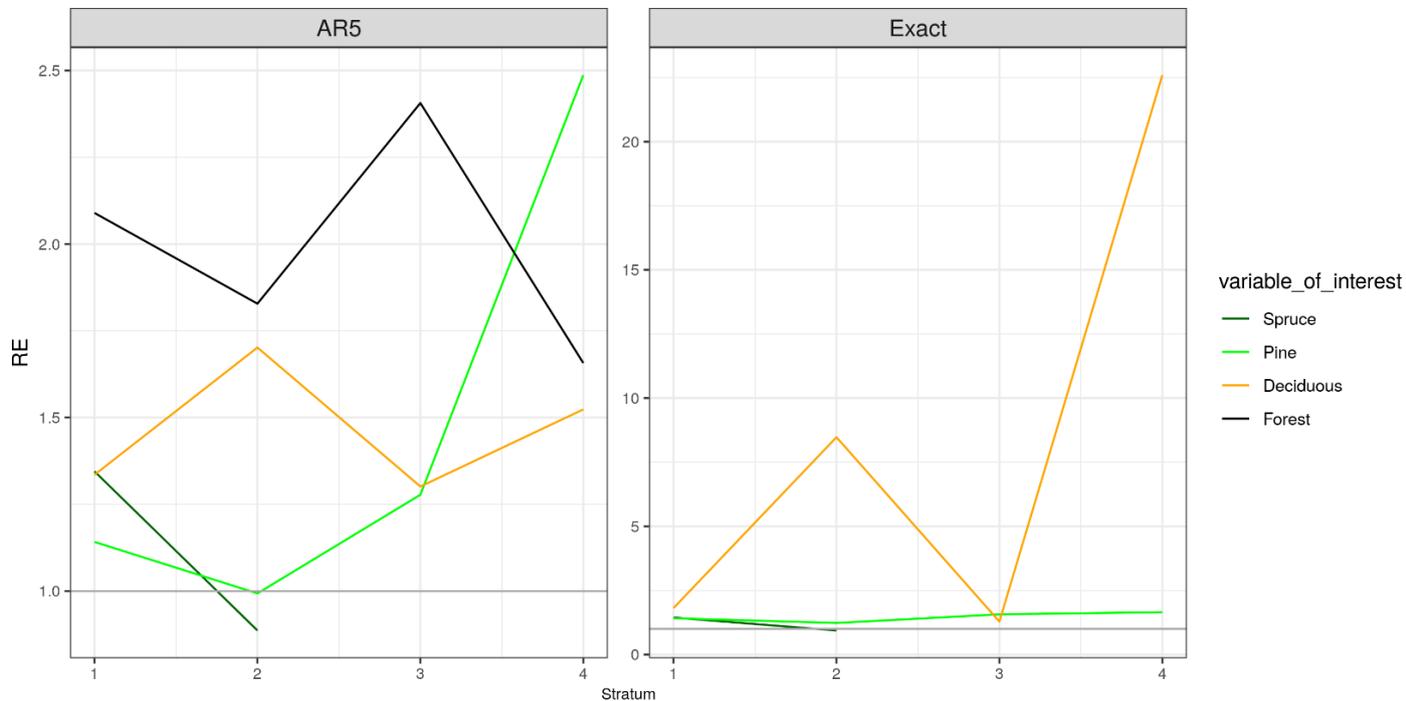

Figure 11: Relative efficiency (RE) of model-assisted area estimates by NFI stratum given the AR5 forest mask and a (theoretical) exact forest mask. Area estimates are forest area dominated by spruce, pine, and deciduous tree species; and forest area. (Stratum 1: lowland forest; Stratum 2: mountain forest; Stratum 3: productive forest in Finnmark; Stratum 4: low-productive forest in Finnmark).

The 355 Norwegian municipalities were used to test the performance of PS and MA estimation for smaller sub-populations. A MA estimate for at least one stratum was possible for 208, 210, 233, and 234 municipalities for spruce forest, pine forest, deciduous forest, and forest, respectively. The remaining municipalities contained less than 30 NFI plots within any stratum. For MA estimates including all strata, the respective numbers of municipalities were 104, 108, 114, and 115. For PS, the respective numbers of municipalities were 71, 60, 34, and 96; and 25, 25, 4, and 22, respectively. In a few municipalities with very high forest cover, PS estimates were possible for spruce and pine forest, but not for forest area because the non-forest group contained too few NFI plots.

The RE of MA estimates of forest area dominated by spruce, pine, and deciduous species, and total forest area within municipalities ranged between 0.2 and 17.6 (Figure 12). Especially with fewer than 10 sample plots within a domain of interest (i.e. spruce forest, pine forest, deciduous forest, or forest area in total), the probability of a RE clearly less than one increased. In the same time, the RE also tended to be highest for municipalities with few sample plots within a domain of interest. With an increasing number of sample plots within a domain of interest, the RE tended to approximate the RE on national level. The largest chance for a clear improvement of the MA estimate over the direct estimate was for municipalities with 10 to 40 sample plots within a domain of interest. However, RE greater than 8 were possible for pine forest and forest area for some municipalities with more than 50 sample plots within the respective domain of interest. Nonetheless, RE slightly less than one were also possible for municipalities with more than 50 sample plots within a domain of interest (Figure 12). The RE of PS estimates for municipalities in which estimates for all strata were possible, were on average between 0.3 and 0.5 units larger than the RE of MA estimates and were never less than one.

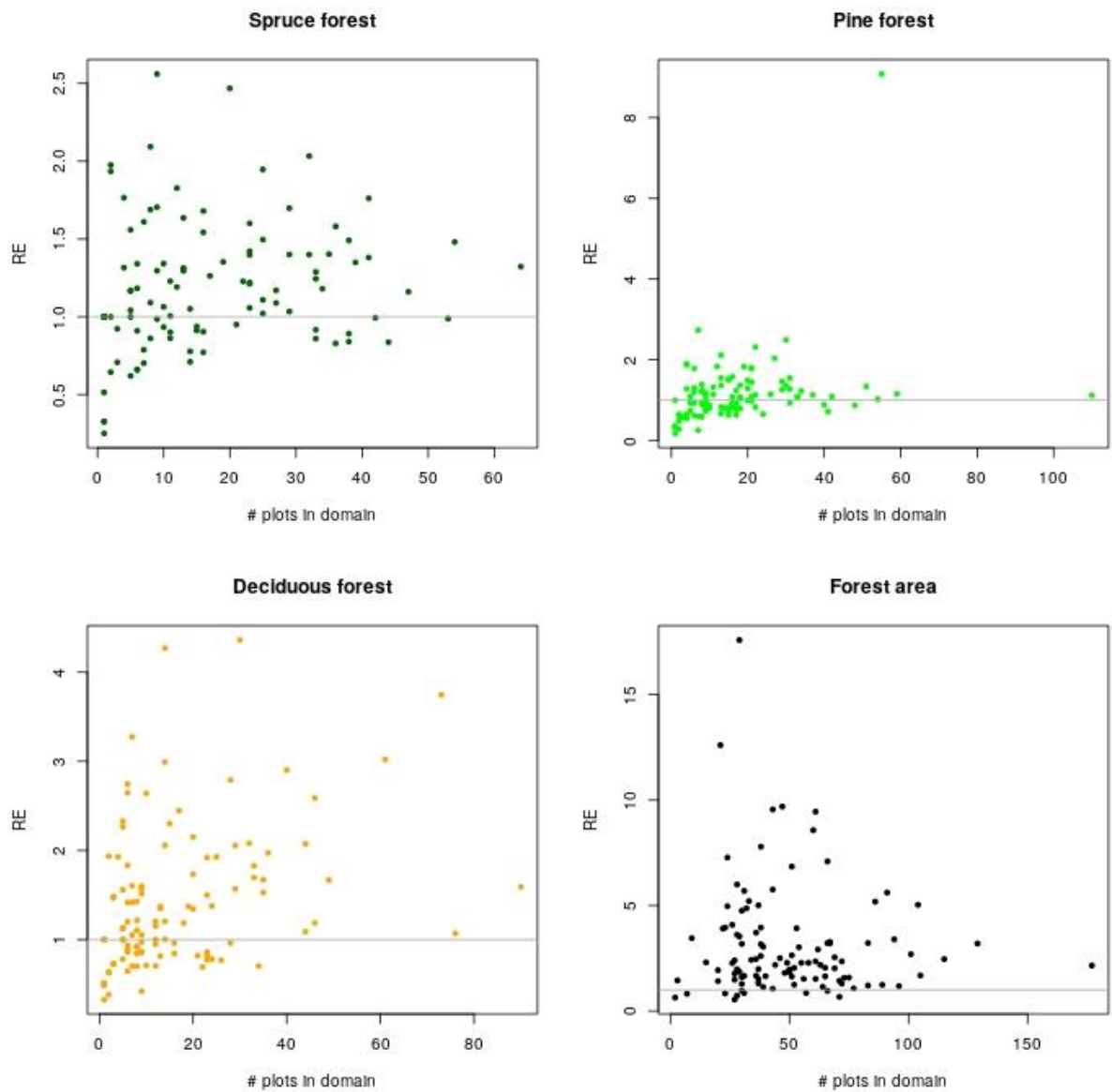

*Figure 12: Relative efficiency (RE) of model-assisted area estimates per municipality given the number of NFI plots over all strata per domain of interest. Area estimates are for forest dominated by spruce, pine, and deciduous tree species; and total forest area. The horizontal line indicates a RE of 1.*

## 5 Discussion

So far, most research on utilizing multi-temporal Sentinel-2 data for tree species classification has focused on case studies for relatively small areas. Still, the obtained accuracies in the present study are in line with these other studies for comparable tree species. For example, Immitzer et al. (2016) obtained 64% OA for six tree species, amongst others for spruce (UA 73%) and pine (UA 34%) for two study areas in Germany. In a follow-up study, Immitzer et al. (2019) showed that further improvements were possible by utilizing a time series of up to 20 inter-annual satellite images. Wessel et al. (2018) classified coniferous, beach, oak and other broadleaved trees with OA greater than 80% for two study areas in Germany. Even higher accuracies were reported by Persson et al. (2018) who classified five tree species in southern Sweden, with OA of 87%, and UA of 78% for pine and 92% for spruce and Grabska et al. (2019) who classified nine tree species using stand-level data in Poland and obtained UA of 100% for spruce and 95% for pine. While the accuracy for pine was the same as in our stand-level assessment, the accuracy for spruce was smaller in our case with approximately four times smaller stands than in Poland. All above mentioned studies used Random Forest as the modelling approach as we did and have the common conclusions that further research is needed when applying the approaches to larger extents, preferable to entire countries.

Nation-wide studies that rely on NFI data suffer from temporal differences between the reference data and the satellite images which will result in somewhat smaller accuracies compared to smaller case studies. Furthermore, we used mosaicked Sentinel-2 data, while single images can be used in most smaller case studies. The advantage of image mosaics is that NFI data as a huge set of reference data are available. The disadvantage is that due to color alignment among images and the combination of images that were not taken simultaneously, the information content in the data decreases which can reduce the model accuracy. When comparing studies, it also needs to be considered that accuracies are affected by differing characteristics of the studied populations and different designs of the reference data (for example sample plot size).

For the national scale, we confirm the findings of several studies on smaller scales (e.g. Persson et al. 2018, Immitzer et al. 2019), that multitemporal satellite data improve tree species classifications. While we used satellite images from only two dates, with an increasing number of species to classify, the number of images from different points of times that improve a classification is likely to increase because of the phenological dynamics of the different species. For national-scale applications as ours, locational predictor variables (i.e., coordinates) can help to describe the phenological dynamics of one species in different parts of the country. Especially in countries with large gradients in the growing season length like Norway, this may be of importance.

The Landsat sensor has, despite a somewhat coarser temporal and spatial resolution, a similar range of applications as Sentinel-2. Because Landsat has one of the longest records of consistent earth observation data, it has been used for tree species and forest type classification since the 1980s (e.g. Walsh 1980, Wolter et al. 1995, Gudex-Cross et al. 2017, Adams et al. 2020). Landsat data also played an important role for forest mapping in the Nordic countries, especially by using the k-nearest neighbor (kNN) approach (Tomppo 1991, Reese et al. 2002, Tomppo et al. 2008). Tree species were in these studies usually mapped indirectly, for example as volume by tree species groups (Tomppo et al. 2008). Following the track of Tomppo (1991), Gjertsen (2007) used a Landsat image with NFI data in south-eastern Norway and reported PA of 75%, 64%, and 33%, for spruce, pine, and deciduous forest, respectively. These values were slightly larger for spruce, but smaller for pine and deciduous forest when compared to our accuracy measures that included errors from the forest mask (county 30). The combined use of Landsat and Sentinel-2 data (Claverie et al. 2018) opens up for further improvements by increasing the number of cloud-free images and the utilization of a long time series.

More recent efforts on fine-resolution large-scale mapping of forest characteristics has been focusing on using 3D remotely sensed data such as airborne laser scanning or image matching (e.g. Nord-Larsen and Schumacher 2012, Nilsson et al. 2017, Astrup et al. 2019), and less on species mapping. Exceptions are Waser et al. (2017) who reported OA of 99% for the classification of coniferous and deciduous forests for the whole of Switzerland using digital aerial images and auxiliary data and Astrup et al. (2019) who reported PA of 66%, 74%, and 59% for spruce, pine and deciduous tree species for a large study area in Mid-Norway using NFI data in combination with image matching data and the associated multispectral information. Similar accuracies in the latter study compared to ours including uncertainties from the forest mask (county 50) suggest that the higher spectral resolution and consistency of optical satellite data and the possibility to use multitemporal data can balance the high spatial resolution of aerial images.

Although the OA of the utilized forest mask was greater than 90%, PA and UA around 70% in the low-productive regions of Norway (NFI strata 2 and 4) were responsible for considerable reductions in estimation efficiency of forest area by tree species. This emphasizes the necessity of a precise forest mask. As an alternative to the AR5 forest mask, Hansen's tree cover 2000 map version 1.6 was considered (Hansen et al. 2013). However, the RE of the MA forest area estimate using the best threshold of a tree cover of 50% was just 1.07 and RE in several strata were less than one. A promising approach to improve the forest mask based on deep learning algorithms was described by Debella-Gilo et al. (2020), but is not yet available nation-wide and was therefore not used.

Improvements of national-scale estimates using remotely-sensed data are rarely reported. However, similar observations in the present study were made for example by Haakana et al. (2020) on the level of municipalities. They used poststratification based on the Finnish multisource national forest inventory to estimate forest characteristics for municipalities and report clear improvements compared to direct estimates. We observed RE <1 for model-assisted estimates in some municipalities which result from cases where models and the resulting maps do not fit well locally. This highlights the importance of poststratification and model-assisted estimation (McRoberts et al. 2016), or other approaches (Tomppo et al. 2008) to reduce local map errors that otherwise can result in biased estimators (Gjertsen 2007). While poststratification was more efficient than model-assisted estimation (McRoberts et al. 2016), the requirement of more sampled observations often did not permit its use, even on the national level.

The tree species map based on the best model is freely available online (NIBIO 2020) as part of the Norwegian forest resource map SR16 (Astrup et al. 2019). The high OA of 90% on stand level suggests that the map may be beneficial in a wide range of applications including forest management planning where species classification is an important part and currently is a largely manual task. Furthermore, models for timber volume or biomass based on airborne laser scanning and image matching improve (e.g. Breidenbach et al. 2008, Schumacher et al. 2019), if they are stratified by tree species. We envision that future research will focus on methods the result in continuously updated maps that improve as new satellite images from a variety of sensors become available and are the basis for improved projections of forest dynamics in the short and long term.

# 6 Conclusion

The following conclusions can be drawn from this study. i) Sentinel-2 mosaics in combination with information on location (longitude, latitude, elevation) were useful for modelling and subsequent mapping of the three main tree species groups in Norway using NFI data as a reference. A forest type map further improved the results. ii) The model accuracy depended on forest structure and was greatest in homogeneous old forests and smallest in young forest or along forest borders. iii) Misclassifications were more likely to occur between spruce and pine forest than between coniferous and deciduous forest. iv) National estimates of forest area dominated by spruce, pine and deciduous species, and forest area as a total, improved when utilizing the prediction map. v) Possible improvements for estimates of smaller areas (municipalities) with fewer NFI sample plots were greater than on the national level. However, direct NFI-based estimates were in some cases more precise than if the species map was used in addition. vi) Poststratification was more efficient than model-assisted estimation but also more restricted in applicability due to the need of more sample plots. v) Further improvements of estimates may be achieved with improved forest cover maps, especially in low-productive mountain regions.

# 7 Acknowledgements

This paper was supported by the Norwegian Institute of Bioeconomy Research and ERA-GAS INVENT (NRC number 276398).

[To be completed after review.]

# 9   Appendix

Table 10: Confusion matrix for model III assuming an exact forest mask and including NFI sample plots not used for modelling (all values in per-cent (%) of the Norwegian land area as represented by the NFI sample plots).

|  |  | Observed | | | | User's accuracy |
|---|---|---|---|---|---|---|
|  |  | Spruce | Pine | Deciduous | Non-forest |  |
| Predicted | Spruce | 7.6 | 1.6 | 1.2 | 0.0 | 73.3 |
|  | Pine | 2.0 | 8.4 | 1.4 | 0.1 | 70.8 |
|  | Deciduous | 1.0 | 1.1 | 13.4 | 0.1 | 85.8 |
|  | Non-forest | 0.0 | 0.0 | 0.0 | 62.3 | 100.0 |
|  | Producer's accuracy | 71.8 | 76.0 | 83.9 | 99.7 |  |
|  | Overall accuracy |  |  |  |  | 91.6 |

Table 11: Accuracy measures and number of NFI sample plots (n plots) by county. (Due to its small size, Oslo county was merged with the surrounding county #30, Viken). See Figure 13 for location and identifiers of the counties.

| County ID | n plots | OA | PA spruce | PA pine | PA decid. | PA non-f. | UA spruce | UA pine | UA decid. | UA non-f. |
|---|---|---|---|---|---|---|---|---|---|---|
| 42 | 1425 | 75.3 | 45.8 | 75.5 | 48.9 | 90.4 | 62.8 | 68.8 | 61.4 | 84.8 |
| 11 | 711 | 85.8 | 51.9 | 69.0 | 64.1 | 94.0 | 73.7 | 68.3 | 69.0 | 91.7 |
| 38 | 1547 | 77.5 | 67.3 | 76.6 | 53.0 | 90.5 | 72.5 | 62.6 | 68.9 | 89.2 |
| 30 | 2371 | 77.4 | 71.4 | 71.7 | 43.8 | 92.8 | 69.2 | 63.6 | 63.2 | 92.5 |
| 46 | 2314 | 88.2 | 47.3 | 74.3 | 71.9 | 94.7 | 65.0 | 71.7 | 70.6 | 94.4 |
| 34 | 4286 | 82.1 | 77.3 | 70.2 | 60.6 | 93.5 | 74.1 | 67.1 | 72.3 | 93.2 |
| 15 | 1001 | 87.0 | 50.9 | 57.1 | 79.4 | 94.3 | 59.2 | 57.1 | 72.2 | 95.9 |
| 50 | 3352 | 81.1 | 73.6 | 59.1 | 58.1 | 91.7 | 70.8 | 55.7 | 64.0 | 92.4 |
| 18 | 2528 | 87.2 | 62.1 | 50.6 | 76.2 | 94.4 | 67.3 | 54.9 | 76.1 | 93.5 |
| 54 | 2473 | 89.8 | 21.7 | 70.5 | 86.6 | 91.5 | 100 | 81.4 | 73.0 | 96.1 |

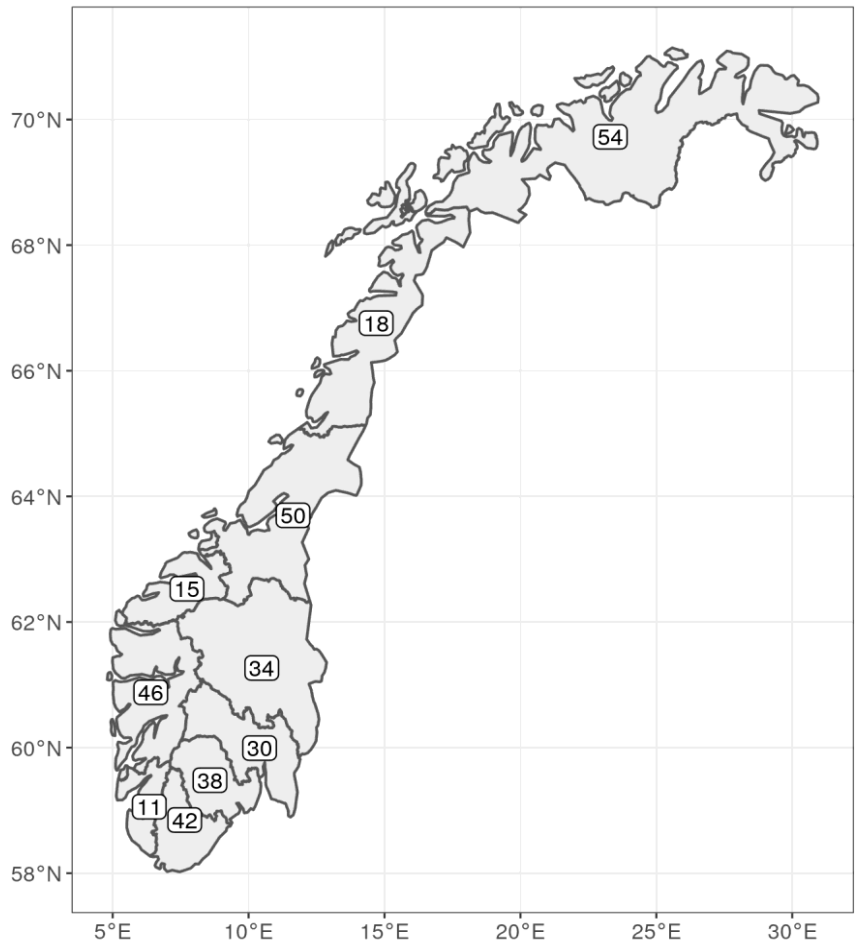

Figure 13: Identifiers and location of Norway's counties. (Due to its small size, Oslo county was merged with the surrounding county #30, Viken)